# Training in translation tools and technologies: Findings of the EMT survey 2023

Andrew Rothwell, Swansea University
Joss Moorkens, Dublin City University
Tomáš Svoboda, Charles University, Prague

**Abstract**: This article reports on the third iteration of a survey of computerized tools and technologies taught as part of postgraduate translation training programmes. While the survey was carried out under the aegis of the EMT Network, more than half of responses are from outside that network. The results show the responsiveness of programmes to innovations in translation technology, with increased compulsory inclusion of machine translation, post-editing, and quality evaluation, and a rapid response to the release of generative tools. The flexibility required during the Covid-19 pandemic has also led to some lasting changes to programmes. While the range of tools being taught has continued to expand, programmes seem to be consolidating their core offering around cloud-based software with cost-free academic access. There has also been an increase in the embedding of professional contexts and workflows associated with translation technology. Generic file management and data security skills have increased in perceived importance, and legal and ethical issues related to translation data have also become more prominent. In terms of course delivery the shift away from conventional labs identified in EMT2017 has accelerated markedly, no doubt partly driven by the pandemic, accompanied by a dramatic expansion in the use of students' personal devices.

## 1. Introduction

Technology is an indivisible part of virtually all translation production and has been for some time. The ubiquity of translation technology has led to it being a key element of translation education programmes. The increasingly dynamic nature of translation technology, particularly since the advent of machine learning and neural networks, has meant that training programmes have had to work hard to keep up, in order to best forearm graduates with the skills demanded by the industry. Despite trainers' best efforts, successive feedback from the industry reports a perceived skills gap in translator training, listing translation technology as the top graduate skill that needs to be improved (ELIS Research 2024).

In order to share best practices, translator trainers have set up networks for information sharing, interacted regularly with the industry, and drawn up frameworks of translator competence as a basis for higher-level university programmes. The European Masters in Translation (EMT) Network enables all three of these activities. The EMT Network began in 2009 as an outward-facing quality label and academic forum to bring together Europe's best master's-level translator training courses with a focus on the dissemination of good practice in order to respond to the needs of industry and international institutions. The sponsorship of the European Commission's Directorate-General for Translation (DGT) was fitting as multilingualism is a basic tenet of the European Union, and the DGT has increasingly built its own technologies.

A project that sprang from the EMT Network, OPTIMALE, carried out a survey of the training in tools and technologies provided by master's level programmes across Europe, the results of which were reported by Rothwell and Svoboda (2012). The technology working group of the EMT Network decided to rerun this survey every five years or so, in order to build longitudinal data to track the changes in technology teaching over time. The second survey was published by Rothwell and Svoboda (2019), and the current study is the latest iteration, looking to answer the following research question and subquestions:

What is the state of the art for master's level translation technology teaching, as represented by survey respondents?

a) How has this changed over the past ten years?
b) How do respondents see this changing in the short- to medium-term?

The aim of much translation technology teaching is to empower graduates, so that they can not only use technologies appropriately, but also build "an awareness of human skills as a differentiator in a technologized employment market, where linguistic, critical, and ethical competences can combine to produce a transversal skill set to equip graduates for the future" (European Commission 2022, 2). This is arguably more important than ever in the light of generative tools based on large language models (LLMs) being made widely available. The survey sought to establish how far, and by what means, master's training programmes have responded to these complementary objectives over the period in question. The intention, as per the main research question, is to offer a snapshot of what teachers are doing, with results producing helpful guidance for others responsible for developing or updating programmes. It was created using Google Forms and open from March 27th to May 23rd, 2023. Invitations were shared within the EMT Network, with other international and national networks, and via emails to programme directors globally, with an intention to solicit responses from programmes worldwide. The full list of questions appears in an appendix at the end of this document. In the following sections, we present a review of relevant literature, followed by a selection of the survey results.

## 2. Literature review

The literature review in Rothwell and Svoboda (2019) provides a historic perspective of translation tools teaching evolving from isolated attempts to a wider practice as well as references to literature summarising relevant sources. Publications on approaches and specificities as well as variations of dedicated courses and course design were mentioned along with those on automation and its impacts on the future of the profession. The emerging translator skills had been covered together with the perceived need by translator trainers to update their curricula according to the dynamic developments in the field.

The current review covers subsequent publications. Education policies outlined in Abdel Latif (2020) represent meta-research on various areas of translation/interpreting learning and teaching practices research, including classroom practices and trainer education-related issues. Broader circumstances of translation pedagogy are also tackled in Kenny (2019) which emphasizes the importance of connecting translator training with technology and the need to eliminate barriers between the two areas, a process for which, we believe, the present study provides extensive evidence.

Most articles and sources surveyed deal with the topic at hand from the teacher/trainer perspective. Course content, approaches, and practices when teaching CAT tools, machine translation (MT), corpora, and localisation are the focus of Giampieri (2021), Zhang and Viera (2021), Shuttleworth (2022), along with Mikhailov (2022) and del Mar Sánchez, Torres del Rey, and Morado Vázquez (2022), respectively. According to Cavalheiro Camargo et al. (2024, 500) translation studies educators have made an "increasing effort to integrate the newest technological advancements into their teaching while still maintaining a critical approach" regarding their use. Our study devotes sections 3.2 and 3.3 to these topics.

Chan and Shuttleworth (2023, p. 269), apart from theoretical contexts and pedagogical approaches, present implementations of translation technology teaching and enumerate "typical content of university courses" in translation technology. Their 12 typical elements comprise terminology, translation memory (TM), MT and post-editing (PE), exploitation of corpora, parallel text acquisition, project management, translation evaluation, customised MT engine construction, localisation tools, audiovisual translation (AVT) tools, speech recognition and speech-to-speech MT, and interpreting technology. A brief comparison between these elements and survey responses appears in section 3.3.

Teaching methodologies are the topic of Toto (2021) and Zappatore (2022). The former considers 'flipped learning' to teach translation technology competences, while the latter focuses on situated translation teaching, using a virtual practical skills laboratory. Rufian (2023) discusses teacher competences in related areas (offering a proposal for a competency model) – cf. section 3.6 below. Rico and Pastor (2022) study "translator educators' beliefs" when it comes to integrating MT into the curriculum. On the latter, see section 3.8 Future Prospects here.

Over the period surveyed here, there seems to have been increased interest in how students respond to technology teaching. Cànovas' (2022) retrospective presents findings of students' perspectives on translation technology, its developments and utilization. The research suggests that students consider translation technology and skills connected with it essential for the profession. More specifically, attitudes towards MT were studied in several instances. In their experiment on PE and revision Scansani et al. (2019, 77) focussed on students' trust in MT, finding "[n]o evidence of a lack of trust". Contrary to that, Torres-Simón and Pym (2021) find that, overall, students remain sceptical towards PE, even though they do recognize its efficiency. The article presents a somewhat negative stance towards MT within the learning community, although as we shall see in the following section, MT now forms a core part of many translation programmes.

## 3. EMT2023 Survey Results

In the following analysis, the results of EMT2023 will be compared, where possible, to those of previous iterations, using the abbreviations OPT2012 and EMT2017. In 2012, the EMT Network comprised 34 members, expanded to 62 in 2014 and 81 in 2019, although the number fell to 63 with the departure of former UK members following Brexit. The network returned to 81 programmes in 2024, with the continued exclusion of UK and Swiss programmes.
OPT2012 received 50 responses and EMT2017 55: 45 from Network members and 10 from institutions outside the EMT. Although EMT2023 participants from institutions with more than one MA programme were asked to complete a separate response for each programme, only two in fact did this, so the 62 responses received come from 60 institutions in the countries shown in Figure 1. For purposes of comparison, numerical results from 2017 will be shown between {curly brackets} and, where given, those from 2012 between [square brackets]. Like EMT2017, EMT2023 was constructed so that completion of all the box-ticking questions was mandatory. Many of the free-text questions that followed them were, however, optional (see the question list in the appendix), so where answers to these are discussed, the number of respondents will be given. Each of the sub-headings that follow corresponds to a section of the survey.

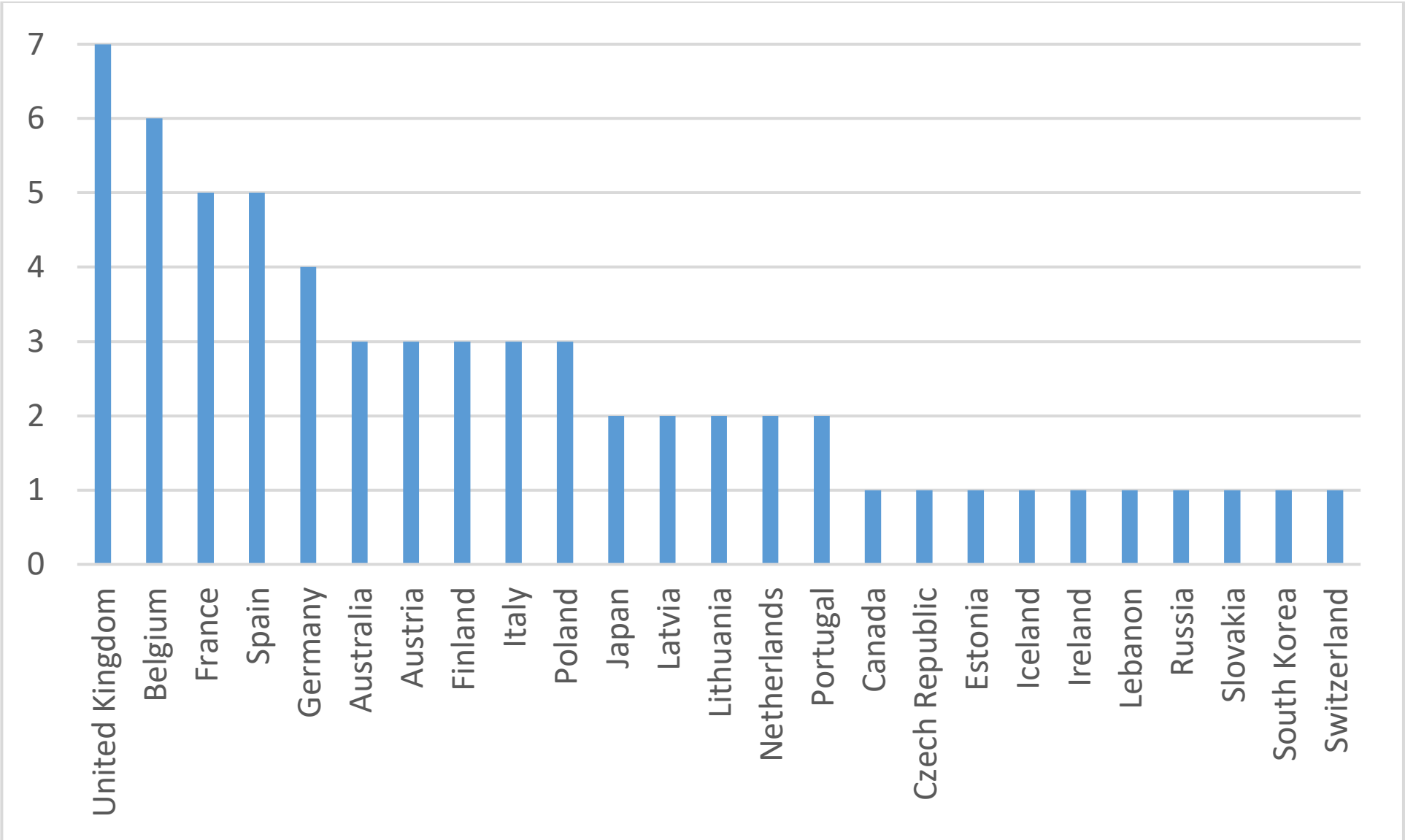

**Figure 1. Responses by country**

### 3.1 Master's Programme Details

Section 1 of the survey captured basic data about programme demographics.
*Q1.4 (network memberships)* showed that of the 62 responses, 60% or 39 came from EMT member programmes, compared to {80% or 45} in 2017. A new question for 2023 asked about membership of other networks: only 20 respondents gave a zero answer, although a small number either did not know or were unsure which might be relevant. Most widely represented was CIUTI with 20 members, followed by associations in France (AFFUMT) and Ireland and the UK (APTIS), with 5 members each. Overall, this result indicates that translation programmes tend to be quite widely networked (one respondent belonged to as many as 6 organisations), arguably a means of ensuring they remain academically current and industry-relevant.
*Q1.5 (ECTS value)* showed that the majority of programmes are still 120 ECTS in size (Figure 2):

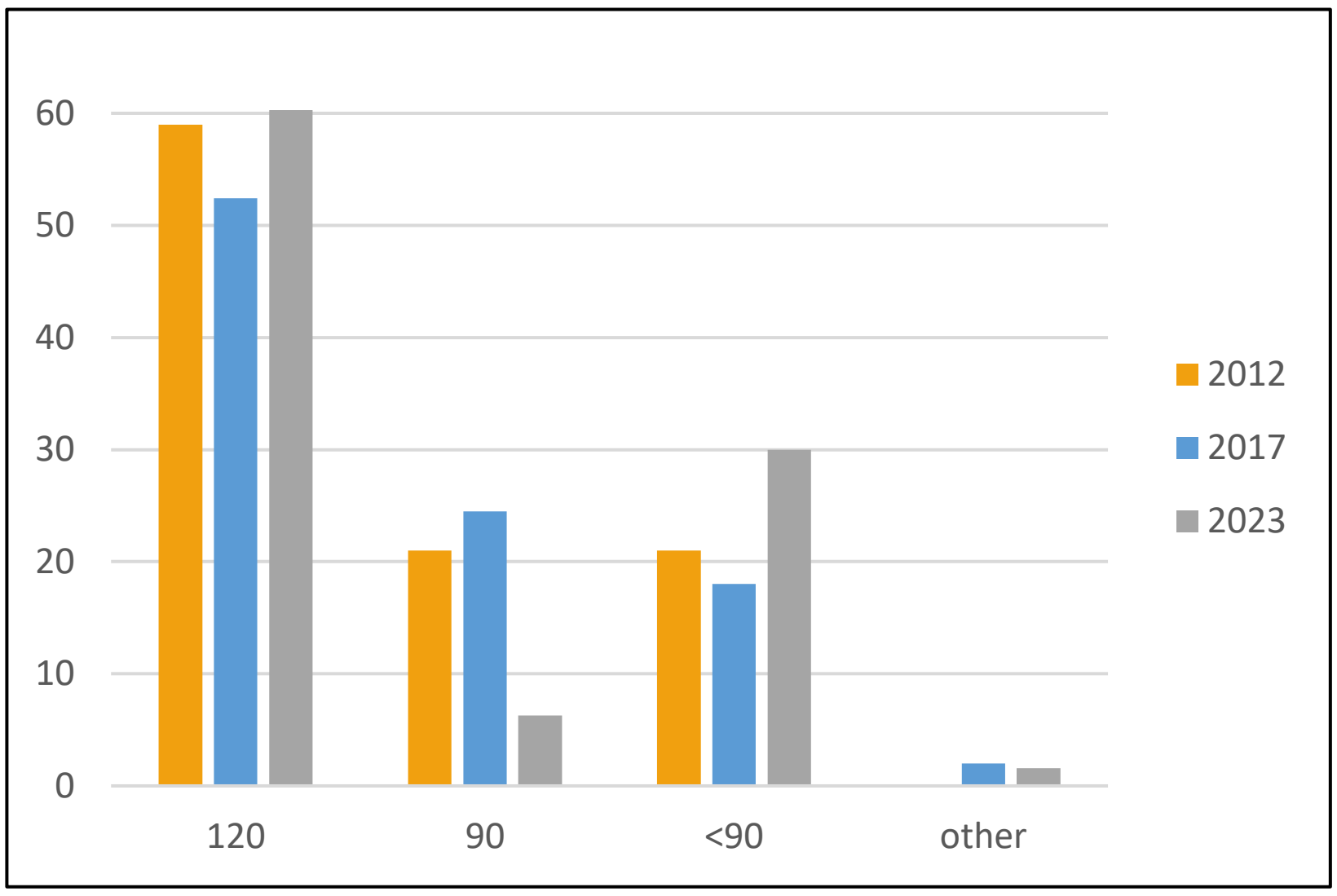

**Figure 2. ECTS values (%)**

The reduction in 90 ECTS programmes is probably explained by the decrease in survey responses (from {14} to 8) from the UK and Ireland. The number of programmes with 60 ECTS increased. Some programmes had variants depending on pathways and language combinations; others had non-standard numbers or did not use ECTS at all.
*Q1.7 (launch year)* again confirmed the major expansion in programmes this century (44 launched since the year 2000). However, whereas the previous surveys showed a big peak in the first decade ({52%} and [60%]) and only two new programmes since OPT2012, EMT2023 identified 24 programmes (39%) launched since 2010.
*Q1.8 (student numbers)* showed a decline in very large programmes (60 students or more), although 6 still reported averaging over 100. At the other end of the scale, 6 programmes (10%) reported fewer than 10 students and 8 between 10 and 19 (Figure 3):

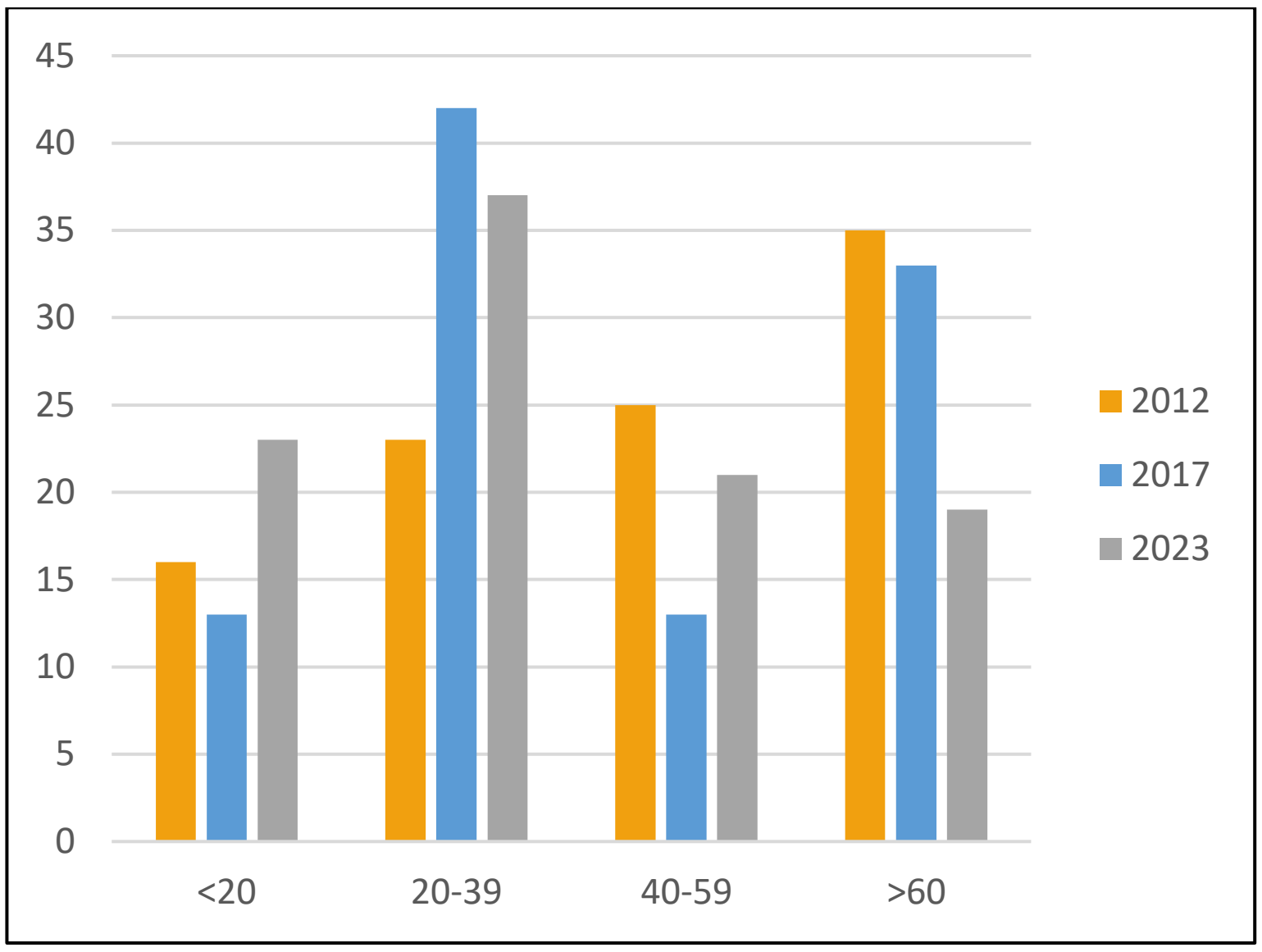

**Figure 3. Student numbers (%)**

This wide variation in programme size, probably related to local circumstances, recent establishment etc., raises questions about organization and comparability of student experience in relation to hands-on tools training, project work, and staffing (e.g. parallel teaching sessions), among other issues.

### 3.2 Approach to Tools Teaching

Section 2 of the survey aimed to establish the broad philosophy of tools and technologies training within MA programmes, and showed a slowing but generally upward trend in the proportion of total study time devoted to technology.

*Q2.1 (teaching and assessment of tools):* whereas in 2017, {100%} of respondent programmes taught the theory and practice of tools (against [93%] in 2012), in EMT2023, eight programmes did not teach the theory, and one programme taught the theory but not the practice. This change may be related to the expanded global scope of the survey.

After a big rise from 2012 to 2017, the estimated proportion of learning time given over to tools (Q2.2 - Q2.9) seems to have declined a little: *(Q2.2)* 42% of programmes had an approximate minimum tools content of less than 10%. The maximum figures (Q2.4) show 42% ({50%} in 2017, [7%] in 2012) allowing over 25% of study time to be devoted to tools, and 6.5% ({9%} in 2017) allow more than 50% (Figure 4):

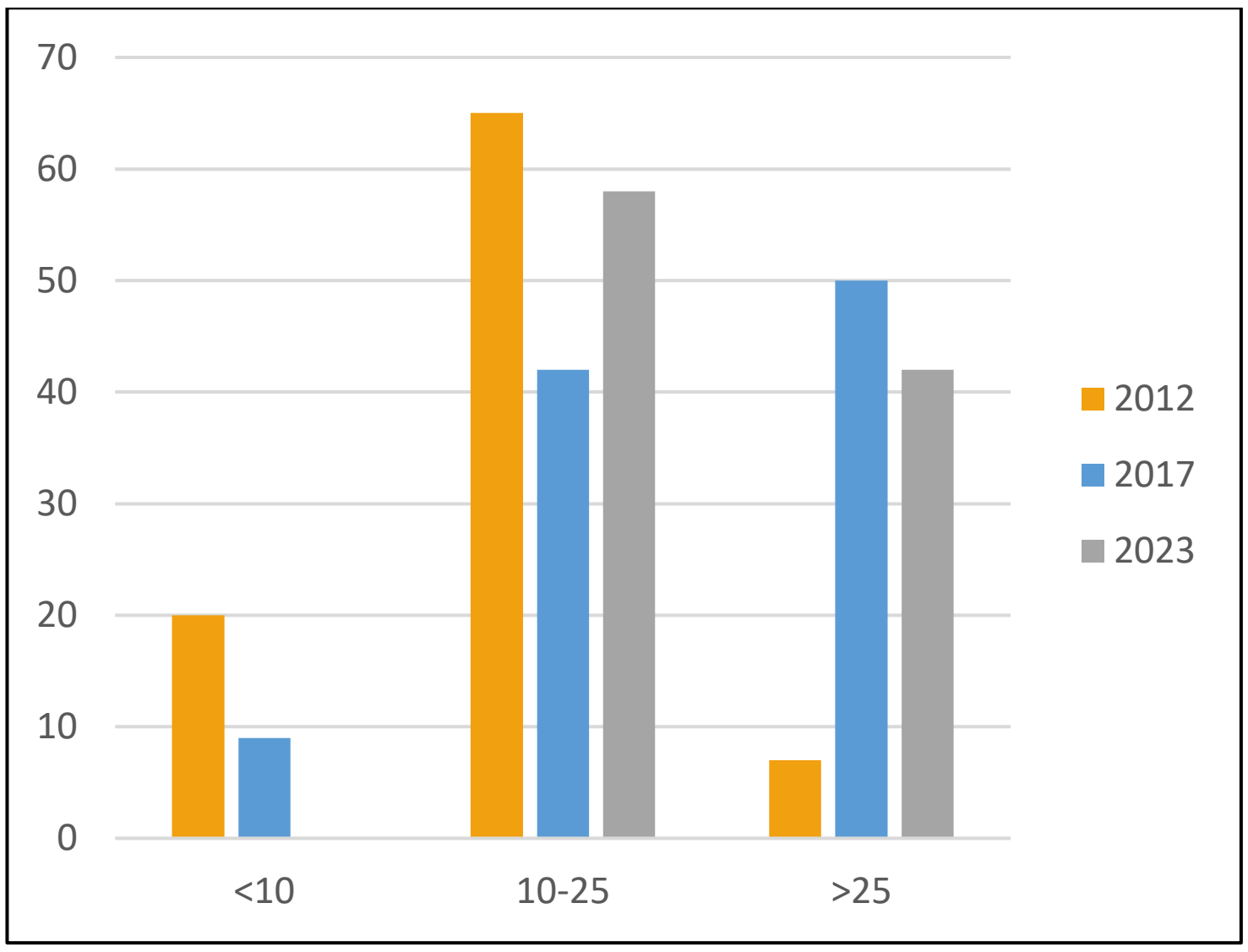

**Figure 4. Maximum study time devoted to tools (%)**

Descriptions of compulsory elements of curricula confirm that tools remain closely integrated with practical translation and professional development modules. Respondents explained that the range of tools-related options, and/or the possibility of writing a dissertation about tools, led to cases where the total tools content could exceed 50%.

In response to *Q2.9 (approach to tools training)*, 85% {89%} [79%] of programmes reported that they taught cost-free tools and 97% {98%} [82%] commercial tools. Virtually all programmes in all three surveys taught tools use from the translator's perspective, but there was a slight decline, after the previous large increase, in teaching from the perspective of the project manager (PM) (79%, against {84%} and [49%]), and also in the use of multilingual projects (89%, against {91%} and [68%]) (Figure 5):

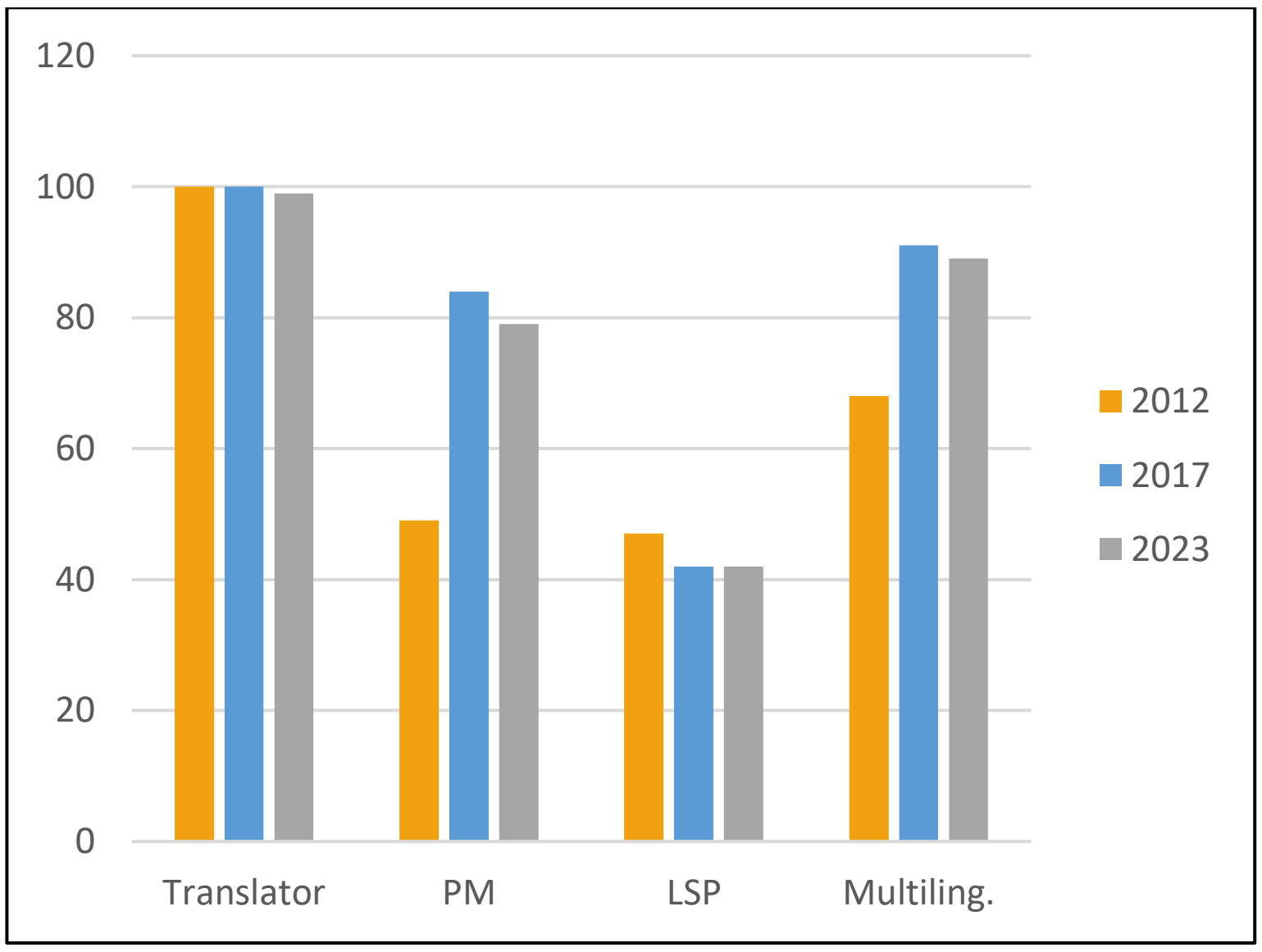

**Figure 5. Role perspectives (%)**

The stability of these results between 2017 and 2023 seems to indicate that, after some quite significant changes from the 2012 survey, professional contexts and workflows remain quite thoroughly embedded across programmes.

## 3.3 Main Activities

*Q.3.1 (training activities)* listed 30 different types of tools use, adjusted from the 2017 list to take into account technological and market developments. It asked whether each was compulsory or optional, and constituted a major or minor part of the programme, or was not taught. In some cases, questions from 2017 were split into several parts relating to PE, MT within/outside a CAT tool, and software localisation (games vs. non-games). Questions about multimedia translation were expanded to differentiate between desktop and cloud-based subtitling tools. A new question on media accessibility was added. Two other new items in EMT2023 reflected technological developments: use of interactive/adaptive MT, and automatic evaluation of MT output using pre-trained metrics (e.g. COMET).

As in the two previous iterations (see Table 1), the top places went to use of TM (89%) and termbases (TB) (87%) - a decline in both figures since 2017, though TM construction by alignment or import comes in at a surprising equal fourth place, perhaps suggesting that programmes are now covering TM technology in greater depth:

| OPT 2012 Compulsory activities | | EMT 2017 Compulsory activities | | EMT 2023 Compulsory activities | |
|---|---|---|---|---|---|
| 1. TM | 70% | 1. TM | 96% | 1. TM use | 89% |
| 2. Termbases | 64% | 2. Termbases | 91% | 2. Termbase use | 87% |
| 3. Data mining | 62% | 3. Data mining | 78% | 3. Info. mining | 85% |
| 4. Project management | 50% | 4. CAT QA tools | 75% | 4. TM construction | 81% |
| 5. Corpora | 40% | 5. PE | 71% | 5. MT evaluation (human) | 81% |
| 6. MT use | 32% | 6. Cloud TM | 69% | 6. PE in CAT | 77% |

**Table 1. Top 6 compulsory activities (tools use – new items in pink)**

In 3rd place, Information mining (split this year between 'search strategies' (3.1.1) and 'evaluation' (3.1.2), with near-identical scores and so combined in the table) increased again to 85%, reinforcing the central role played by research and data-gathering skills. Although PE continued to progress, it was pushed into 6th place by Human evaluation of MT, equal fourth on 81%, a remarkable rise from 8th place and {53%} in 2017. This reflects the increased prominence of MT in the translation workflow since the advent of neural approaches, in their infancy in 2017. Taken together with the figures for PE and use of MT within a CAT tool (6th place, 77%) and outside the CAT environment (7th place, 71% - see Table 2), this suggests very significant expansion in the centrality of MT in today's translator training programmes. In addition, the increased use of automated MT evaluation metrics (string-based and pre-trained), taught by 55% of programmes, when combined with the greatly increased prominence of human evaluation, suggests a more sophisticated and rigorous approach to MT quality. However, at the moment MT usage itself seems to be mainly from the perspective of a user of standard, free online MT: only 31% of programmes reported teaching Interactive/adaptive MT (a new question for 2023), and 16% the training of MT engines (the same figure as in 2012, back in the statistical MT era). It will be interesting to see how these figures evolve in the light of future technological developments and new data-related roles for linguists.

In this perspective it will also be interesting to observe whether, in a future iteration of this survey, conventional CAT tools as represented by the scores for TM and TB use (Table 1) will continue to dominate as they have done over the past decade. The EMT2017 survey identified Cloud TM as a significant new entrant, in 6th place with {69%}, and given the strong development of Cloud-based tools since then, their absence from the 2023 list may seem surprising. However, in the new survey the question was rephrased and divided into 3.1.8 Shared TM and 3.1.9 Shared TB, to capture whether cloud-based tools are being taught in stand-alone or collaborative mode, and the relatively high scores (9th place, 65% and equal 11th place, 60% respectively) suggest that many programmes are taking advantage of the collaborative possibilities these tools offer.

| | | | |
|---|---|---|---|
| 1. TM use | 89% | 11. Corpus construction | 60% |
| 2. Termbase use | 87% | 12. Shared termbases | 60% |
| 3. Information mining | 85% | 13. Translation management systems | 55% |
| 4. TM construction | 81% | 14. Subtitling (desktop) | 40% |
| 5. MT evaluation (human) | 81% | 15. Auto MT eval (string) | 39% |
| 6. PE in CAT | 77% | 16. Website localisation | 39% |
| 7. PE outside CAT | 71% | 17. Standalone QA tools | 37% |
| 8. QA in CAT | 70% | 18. Software localisation | 35% |
| 9. Shared TM | 65% | 19. Interactive MT | 31% |
| 10. Term extraction | 61% | 20. Subtitling (cloud) | 27% |

**Table 2. EMT2023 compulsory activities (tools use) 1-20**

A striking comparative aspect looking longitudinally is the very strong rise in the percentage figures from one survey to the next. In OPT2012, the 7th most prominent activity figured in only [30%] of programmes and the 12th in just [16%]. Whereas in EMT2017 the 12th activity had risen to {35%}, in the latest survey the figure is 60%, which is strongly indicative of the much-increased density of these core technology-related activities (information mining, CAT and MT) across programmes. A consequence of this, compounded by the splitting of some key questions as discussed above, is that a wide range of other technology-based translation activities are widely taught, as may be seen in Table 2. Considering that evaluation of MT output using pre-trained metrics is quite a recent phenomenon and not straightforward to implement, it is not surprising that it is a compulsory element in only ten programmes and optional in nine others. Mapping Chan and Shuttleworth (2023)'s 12 topics as 'typical course content' (see above) onto our topics catalogue we observe substantial overlap. Some of the differences, though, involve the following: (i) our strict focus on translation, (ii) our catalogue differentiates some aspects in more detail (e.g. TM use, TM construction and Shared TM), (iii) we distinguish Information mining and Quality assurance (QA) as additional items and (iv) the former approach features Customized MT Engine Construction and Speech Recognition as specific categories, whereas in our survey, although surveyed as well, these aspects have not made it into the 20 compulsory topics list.

The increase in the diversity of tools-based activities reported in 2017 appears to have continued to 2023. In free text responses, respondents include the use of Jupyter notebooks (as proposed by Krüger (2022) and Moorkens et al. (2024)), introduction to programming for translation and more advanced macro learning. Many report the inclusion of tools (in some cases including LLMs such as ChatGPT) in simulated translation bureaus/skills labs and in practical translation modules, as reported in 2017, though one regrets that "the challenge still remains to encourage colleagues to introduce relevant technologies into updated versions of their courses".

The incorporation of tools into modules not specifically devoted to them was probed further in Q3.3 (compulsory role of tools in other modules). 70% of programmes (up from {55%} in 2017) reported that tools were a compulsory element of practical translation classes and 68% ({65%}) of modules on the translation profession. For 63% of programmes (a significant rise from {45%} in 2017) they are also a compulsory part of translation theory courses, and for 61% ({62%}) of skills lab-type modules. 52% of internships (down from {58%}) have a compulsory requirement for tools to be used, 48% of extended translation project modules (up from {42%}) and 39% ({36%}) of dissertation modules. Apart from increased involvement of tools in practical translation classes, the most notable development here is a wider apparent acceptance that the way tools work should be included in theory modules. Some respondents pointed out in Q3.4 that tools integration into other modules could be lecturer-dependent, as well as subject to student choice.

### 3.4 Tools Taught

Section 4 of the survey aimed to establish which translation tools are most widely taught. In view of the larger number of tools available, the alphabetical listing from 2017 in Q4.1 was subdivided between CAT, terminology and localisation, MT, and AVT. In each case respondents were asked to specify whether a given tool was compulsory or optional, and the approximate ratio of students to licences. In the case of compulsory CAT, terminology and localisation tools the top 6 list (see Table 3) was closely similar to the EMT2017 one, with the same top 4 in the same order (Memsource Cloud having been rebadged as Phrase TMS), and very similar percentages (but see the comment below on the ratio of MT to CAT tools). Trados Studio and memoQ have made significant moves to integrate their desktop offerings with the cloud and fend off competition from the cloud-native new generation (see Rothwell et al. (2023) on these changes).

| **OPT 2012** Compulsory tools | | **EMT 2017** Compulsory tools | | **EMT 2023** Compulsory tools | |
|---|---|---|---|---|---|
| 1. SDL Trados Studio | 56% | 1. SDL Trados Studio | 82% | 1. Trados Studio | 77% |
| 2. SDL MultiTerm | 46% | 2. SDL MultiTerm | 69% | 2. MultiTerm | 68% |
| 3. SDL Trados 2007 | 40% | 3. memoQ | 55% | 3. memoQ | 53% |
| 4. Google Translate | 26% | 4. Memsource Cloud | 44% | 4. Phrase TMS | 47% |
| 5. memoQ | 22% | 5. Wordfast Anywhere | 33% | 5. MateCat | 26% |
| 6. SDL Passolo | 22% | 6. OmegaT | 31% | 6. OmegaT | 18% |
| | | | | 6. Wordfast Anywhere | 18% |

**Table 3. 'Top 6' CAT, terminology and localisation tools**

The freeware OmegaT and formerly (until January 2023) free-to-use Wordfast Anywhere, in equal 6th place with 18%, have both lost significant ground since 2017 despite being cost-effective to teach (Figure 6).

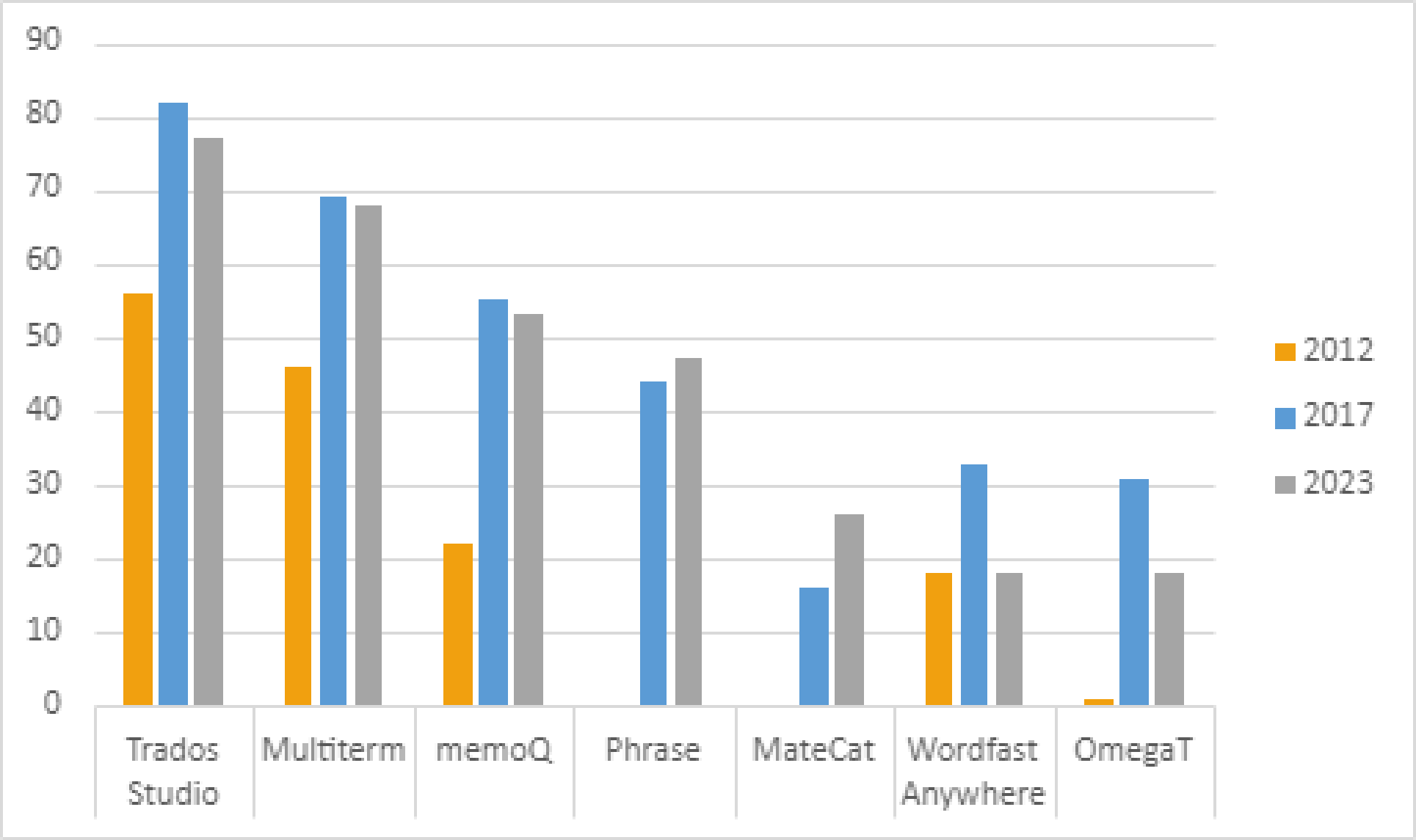

**Figure 6. Top 7 CAT tools, 2012-2023 (%)**

Among other tools, only Passolo (16%, 8th place, down from {29%} in 2017) had compulsory users in double figures (10), while other localisation tools were barely represented. Perhaps surprisingly, the recent licencing changes for Trados Studio, offering free licences for online sign-in, have not yet made an impression on the use of the tool for teaching. Of the programmes that do not teach Trados Studio compulsorily, most teach memoQ and Wordfast as a compulsory tool, although seven programmes listed no compulsory CAT tools at all.

Despite being available as options in some other programmes, none of the other 'traditional' workstation CAT tools were compulsory in more than 3% to 8%. Apart from Matecat, the 'new generation' of cloud-based, MT-integrated CAT tools fared little better: 0-6% use Fluency Now, Lilt, Smartcat, Smartling, XTM, or Wordbee (at 5% down from {15%}). Although a few programmes make a dedicated PM tool compulsory (Plunet 8%, XTRF 10%), it seems likely that the relatively comprehensive PM features offered by cloud CAT tools high on the list fulfils the PM training needs of many. Equally, the first-to-market Matecat seems at the moment to meet the needs of programmes teaching the integration of CAT and MT. Whether there will still be a meaningful distinction to be made between CAT and MT in a few years' time remains, of course, an open question.

For the first time in these surveys, we subdivided CAT, MT and AVT tools. The two most popular MT services are Google Translate, compulsory on 56% of respondents' programmes, and DeepL (55%). If MT tools had been ranked in the same table as CAT and localisation tools, as in 2017, Google Translate and DeepL would have come in 3rd and 4th places respectively, again showing a very significant advance in the role of compulsory MT in translation programmes (in 2017, only MT@EC with {20%} and Kantan MT with {18%} figured in the list of top 12 tools). The European Commission's eTranslation service (successor to MT@EC), the availability of which is limited to EMT member programmes, is now compulsory according to 24% of responses. Microsoft Bing Translator is compulsory in 19% of programmes and Systran in 8%. Possibly due to the discontinuation of their free academic programme, KantanAI (formerly KantanMT) has dropped from 18% in 2017 to just 2% in 2023.

The most-used compulsory AVT tools are the workstation-based Aegisub and Subtitle Edit (both 18%), followed by Subtitle Workshop (16%). EZTitles, Ooona Tools, and the Trados Subtitling App are all compulsory in 8% of programmes. This reflects the importance of AVT to a small group of programmes, with a marked preference for freely available tools, as few software publishers offer free or discounted academic licences.

Free-text responses highlighted as priorities: (local) market relevance of tools, the need to select a small number of 'representative' tools, and student access (ensuring students can access any taught tool on a one-to-one basis as required). A large majority of respondents emphasised budget constraints and the practical requirement to use free tools, trial versions, and academic licences, which, when put together with the need to select a small number of representative tools, suggests that vendors which do not offer low-cost or cost-free solutions to universities are less likely to see their software selected.

*Q4.3 (tools or technologies respondents would like to teach but are unable to):* technological areas that respondents would like to develop further include MT training, MT integration into CAT tools ("a problem, because most MT plugins are commercial and using them would cost a fortune"), PM, accessibility, OCR, speech recognition (currently taught on 13% of respondent programmes), terminology management, and LLMs. Specific applications that respondents mentioned as desirable but unimplemented included EZTitles, Ooona, Systran, XTM, Acrolinx, Kantan, Sketch Engine, memoQ, ChatGPT/GPT-4. The most frequent reason given for not introducing them was budgetary constraints, followed by lack of available class time, and insufficient staff resources and/or expertise.

**Q4.4 (integration of MT into CAT tools)* A substantial majority of respondents (63%) reported that they teach this aspect of translation technology. Among 41 free-text respondents, several reported using API keys paid for by staff to demonstrate the principle of MT integration "while shielding students from the costs involved". Significantly more respondents use free MT integration offerings of CAT tools. Several stressed the interest of using open-source tools such as Opus-CAT/OpusMT (Nieminen 2021), including in OmegaT, though the plugin only supports Windows. 81% of respondents do not use paid versions of MT tools. Among the 18 respondents to a free text question asking for details, several confirmed that paid licences were held by teachers and one said that the programme had a subscription to DeepL.

**Q4.6 (ChatGPT and other text generators)* Of 57 respondents, only 12 reported that they did 'not yet' teach this technology, but all stated or implied that they expected to do so soon. Among the experimental approaches already in place, several programmes compare ChatGPT with conventional MT output. Others are using it in revision, text adaptation for specific purposes, and experiments with prompts for domain-specific output. Finally, several programmes teach the operating principles of LLMs and their relationship to NMT from a developer perspective, and one is considering restructuring the course around generative tools.

## 3.5 Tools Themes

Section 5 of EMT2023 looked at the perceived relative importance of a range of tools-related themes in respondent MA programmes.

*Q5.1 (themes)* listed 15 topics, including the history and development of translation tools, tools within translation projects, professional workflows, PE (modified this time to focus specifically on ergonomics), industry standards, legal issues, and tools in crowd-sourced and volunteer translation, inviting respondents to rank the importance of each, and to suggest further themes not included in the list. Two new questions this year were on ethics of translation technology, and LLMs.

The top two themes (see Figure 7) remain, as in both previous surveys, Tools as part of a translation-related project (82%, up from {75%} and [58%]) and Tools in professional roles and workflows (79%, up from {73%} and [64%]).

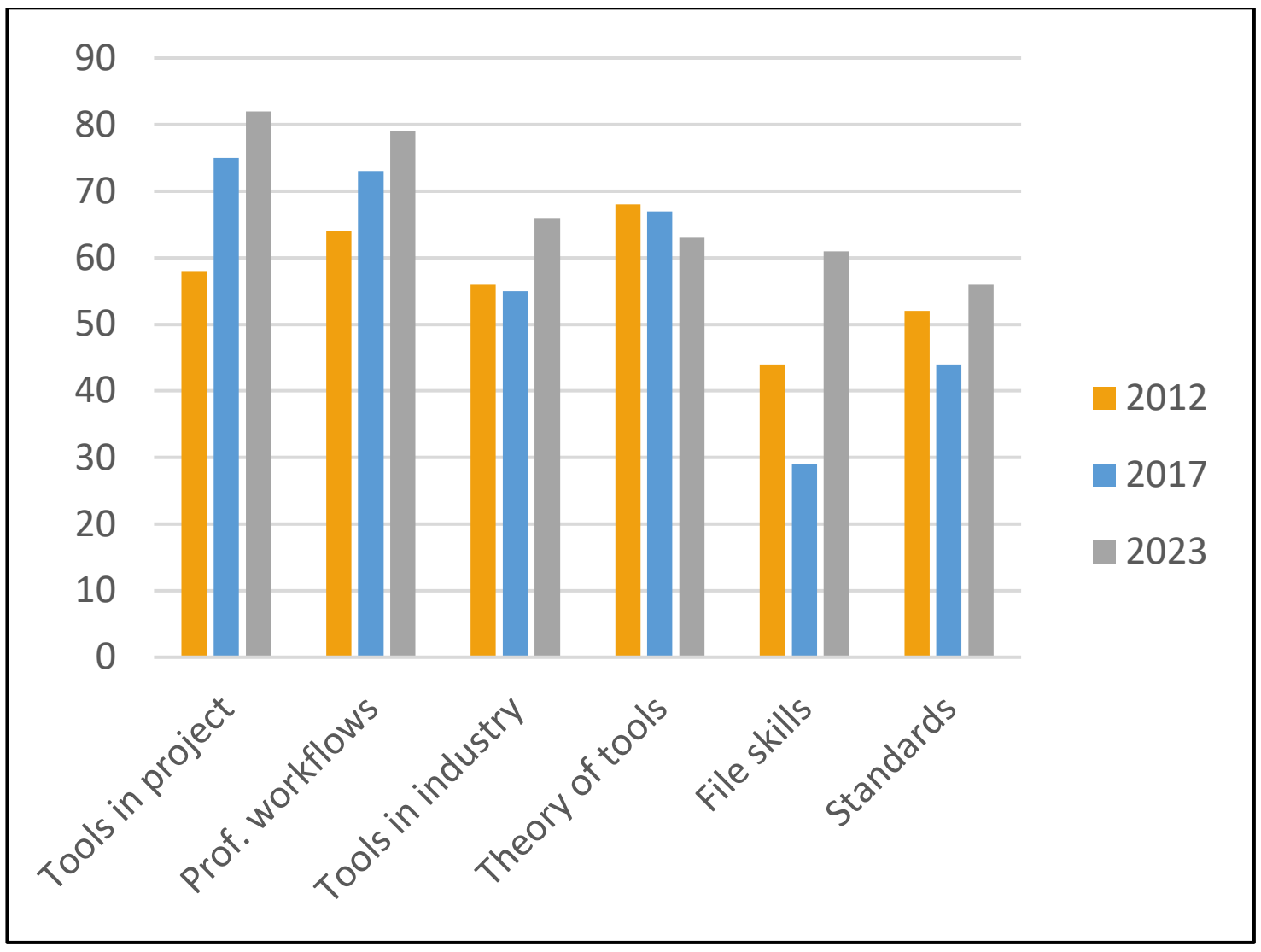

**Figure 7. Top 6 tools themes (%)**

In EMT2023 the place of tools in the translation industry rose from fifth to third place in the list ({55%} to 66%), reversing the decline from 2012 and narrowly pushing the theory of tools from 3rd to 4th place with 63%. In 5th, significantly up from 11th in 2017, comes Generic file management and data security skills (61% as against {29%}). Ethical and legal issues are prominent in the lower half of the list (see Figure 8), perhaps in reflection of the increased importance of issues around NMT/LLM training and data ownership.

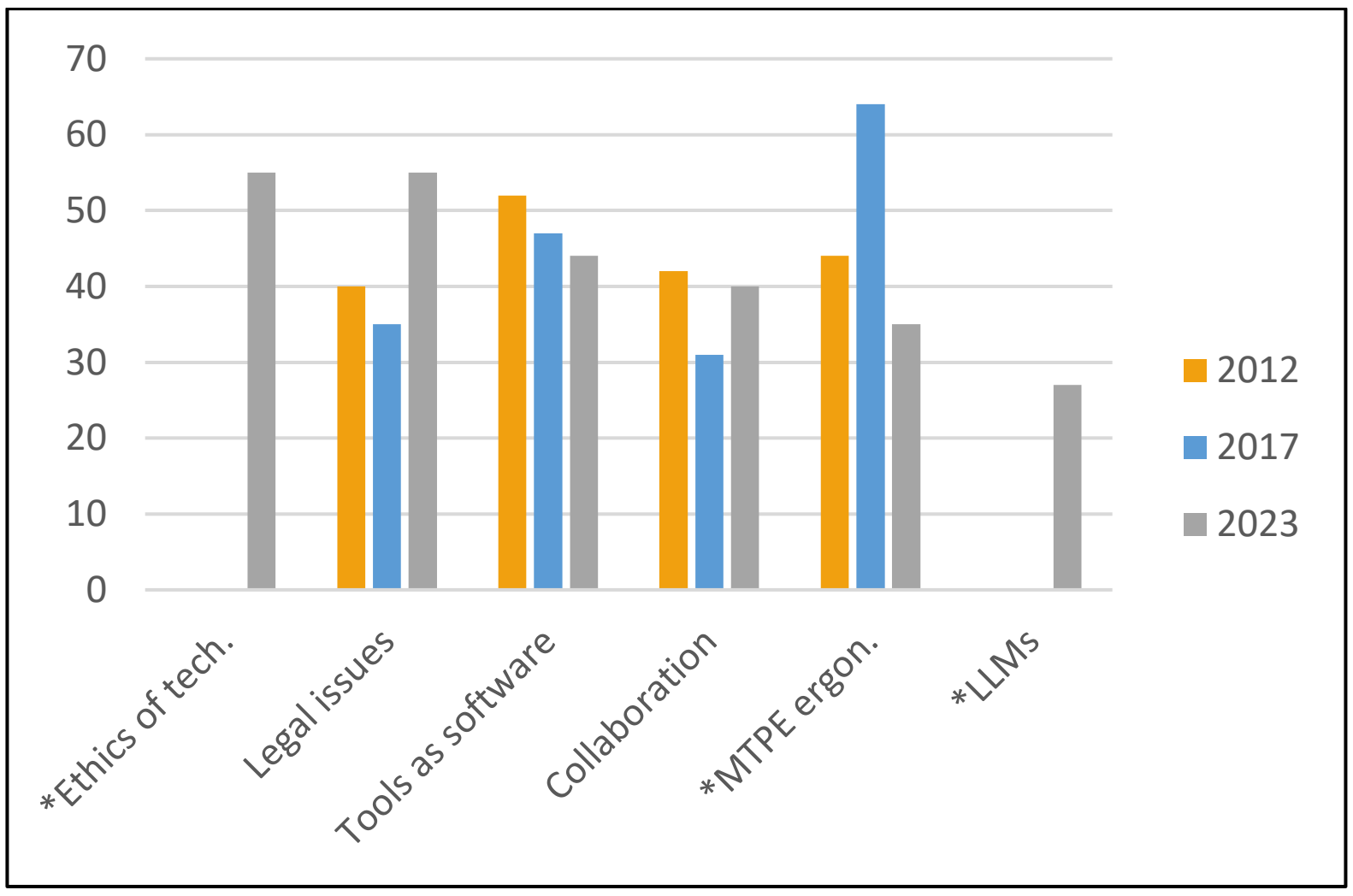

**Figure 8. Tools themes 7-12th place (%)**

The perceived importance of teaching tools with a focus on their attributes as software has continued to decline slowly. The large 2017 spike in the importance of PE issues has dropped substantially, perhaps because the question was narrowed this time to focus on ergonomics, but possibly also because PE is now a better understood and less controversial part of translator training than it was. Finally, the perceived importance of LLMs, which only came to public notice in the months preceding the survey, indicates the responsiveness of programmes to technological innovations, as well as hinting, perhaps, at a future significant concern for translators. Overall, these results show that themes related to translation tools in real-world professional contexts have continued to increase in perceived importance.

This was confirmed by many of the 22 free-text responses to Q5.2, which invited comments on tools themes. Some emphasised practical skills acquisition, usability assessment of tools, research into industry trends, ethics and legal issues, corpus, terminology and workflow management, and QA. Others mentioned XML, web accessibility, HCI, and again, LLMs.

*Q5.3 (teaching and assessment strategies)* asked respondents to identify which of 6 teaching and 6 assessment strategies their programmes used, and to rate their importance on a Likert scale of 1-5. Among teaching strategies, Learning through staff lecture/demonstration (82%) came behind Individual work (92%) and narrowly ahead of Team and group work (81%). It is tempting to relate all the increases in autonomous, blended, and e-learning (Figure 9) to adaptations forced by the Covid pandemic (as suggested by Afolabi and Oludamilola (2022)).

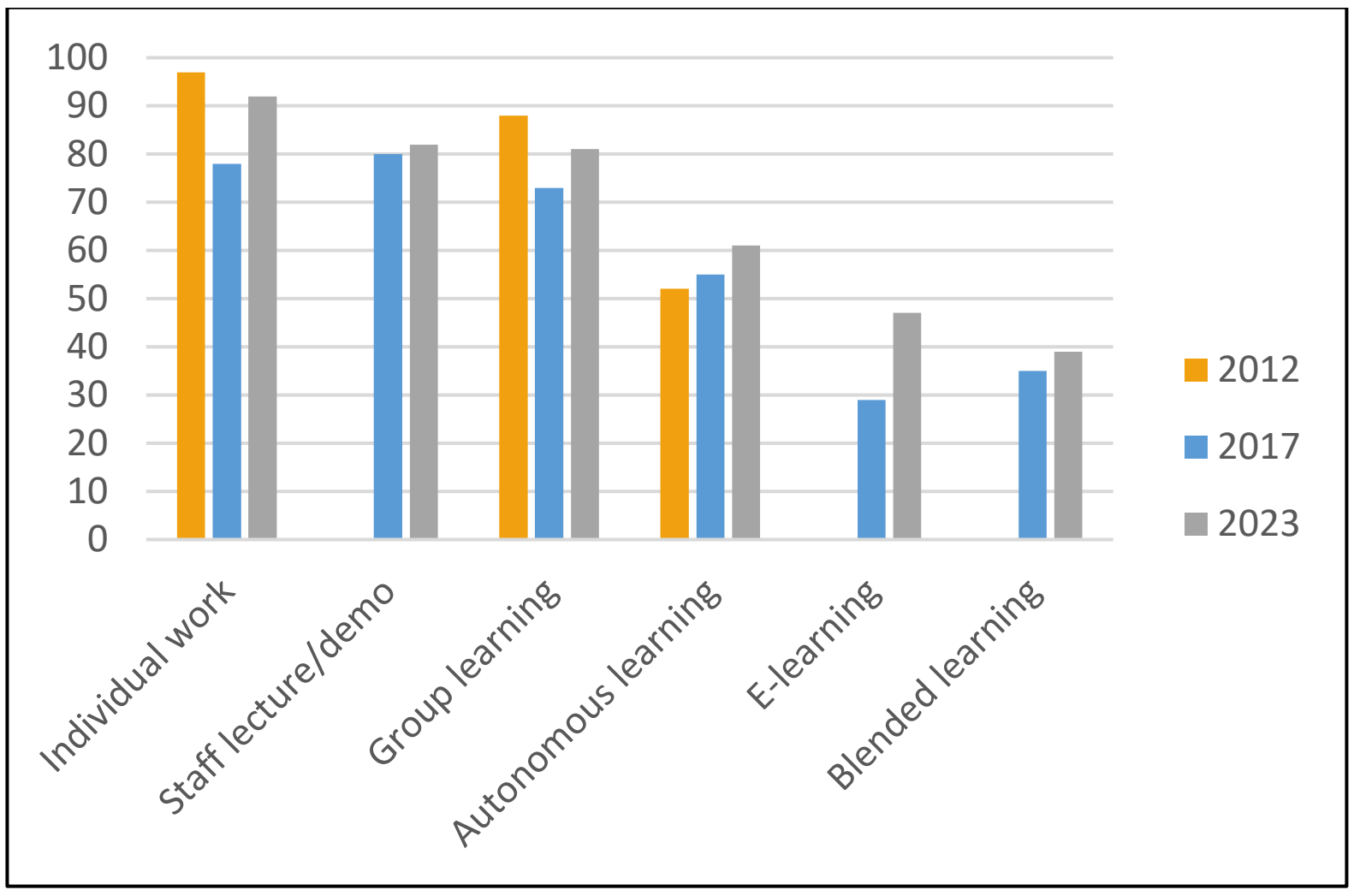

**Figure 9. Teaching strategies regarded as (very) important (%)**

In terms of assessment of students' tools knowledge, Assessment by both individual and Team/group work increased in importance from 2017 (see Figure 10):

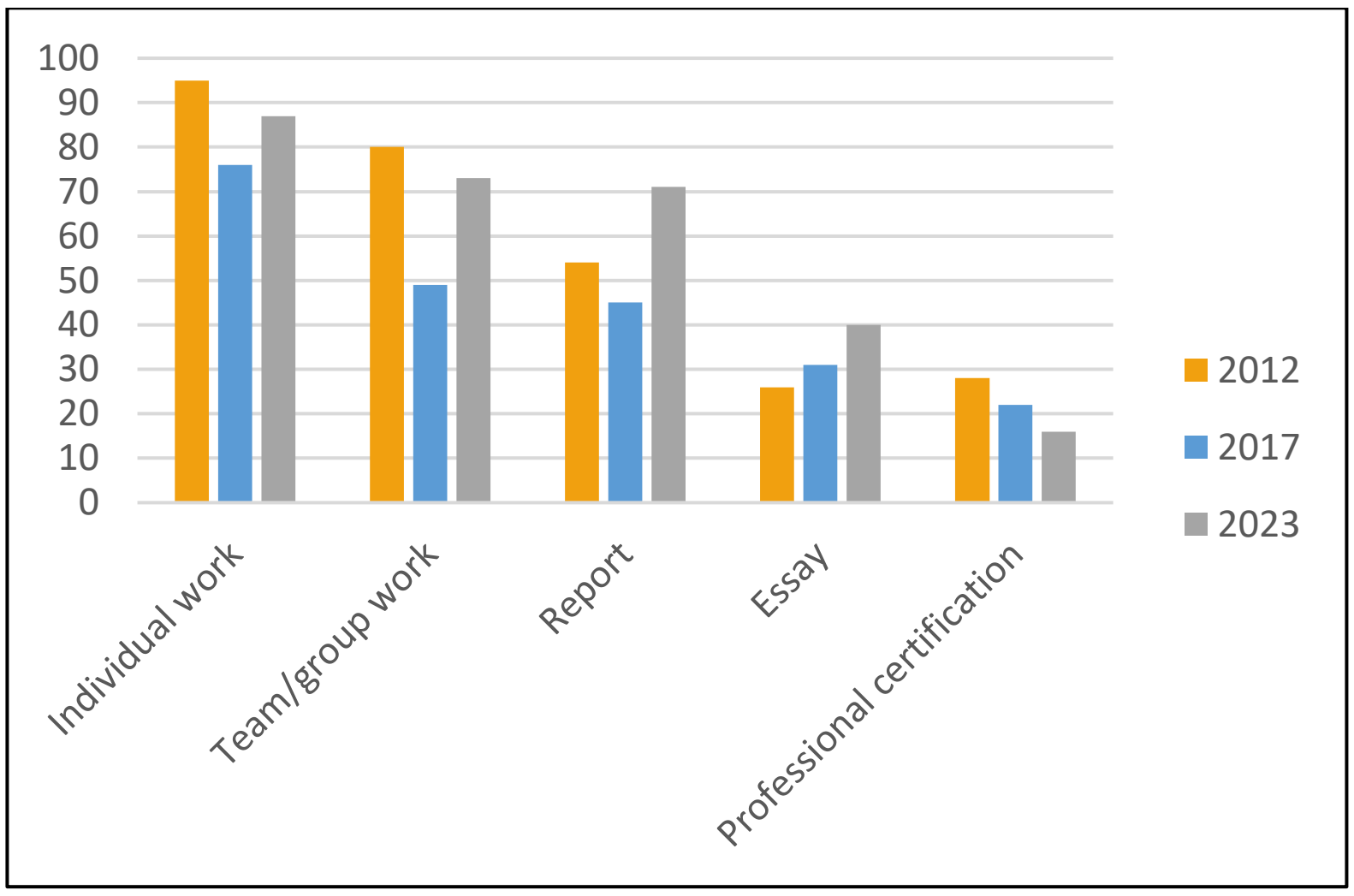

**Figure 10. Assessment strategies regarded as (very) important (%)**

The significant increase in the perceived importance of Reports (71%, up from {45%}) and to a lesser degree Essays (40%, up from {31%}) may also be a side-effect of the pandemic making assessment by practical tasks harder to implement. The decline in the perceived importance of professional certification tests for programme assessment has continued (16%, down from {22%} and [28%]), perhaps suggesting a mismatch between their assessment objectives and methods and those of academic training programmes.

*Q5.4 comments:* several among the 23 free-text comments on teaching and assessment strategies mentioned student-centred, autonomous learning and "authentic assessment", defined by one as “a combination of practical tasks and critical reflection”. This perspective may become more common in response to the prevalence of LLMs. One respondent wrote of “co-construction of learning/assessment needs: asking students to identify skills, tools and criteria for evaluating a given project”.

**Q5.5 (Did the necessity of remote delivery and collaboration during the COVID-19 pandemic have an influence on the i) content ii) delivery iii) assessment of your programme?)* generated 53 substantive text-based responses, of which only one bluntly said ‘No’. One respondent reported little difference in content and delivery, but problems with assessment, while two remarked that they were already using blended learning anyway, so there was little difference. Most respondents reported a shift, at least for particular time periods, to online delivery, but there were a few negative comments about this (one respondent noted a negative effect on group work, which had to be done in person, another that internships had become ‘virtual’). On the contrary, there were quite detailed descriptions of some of the strategies adopted to make a success of online delivery (both synchronous and asynchronous), particularly the development of recorded software demos and videos, and one respondent commented that such developments had enhanced the opportunities for independent learning. There were fewer comments about assessment: one programme had moved to using vendor certification tests as a result of lockdowns and another had successfully moved from an exam to a report format.

**Q5.6 (Have you retained any of these changes post pandemic?)* Among the responses to this follow-up question, 9 were negative but 40 clearly positive, in varying degrees. One translation technology course was reported as being still entirely online, while 6 continue to be delivered in hybrid form. Another retains its use of virtual labs, and several respondents also said that access to improved classroom facilities has continued post-pandemic. Four respondents said the new materials developed to support online learning continue to be used and appreciated by students, and one that remote technical support for students is still delivered via Zoom. Two programmes have continued with a teleworking element in their internships, and several report a greater use of cloud-based software than before the pandemic. Among other benefits to come out of the experience, several respondents reported that staff were now more comfortable with online delivery, flipped classrooms, and autonomous learning, while one said that there had been a change of attitude among teachers to MT, now no longer regarded as simply ‘cheating’.

## 3.6 Teaching Staff, Student Access, IT Support

Section 6 of EMT2023 was devoted to the material conditions of tools training delivery within programmes.

*Responses to Q6.1 (who teaches tools and technologies?)* showed that in 2023 24% of programmes relied solely on salaried academic staff, as against {22%} and [38%], although 37% mostly used such staff (as against {42%} and [25%]), giving similar figures for overall reliance on academics of 61%, {64%} and [63%]. 26% (vs. {22%}) of programmes reported a roughly equal balance between academics and external professionals, 5% {7%} used a majority of external trainers, and 3 programmes (5%) relied entirely on such experts, making a total of 36% of programmes that depended for half or more of their teaching on external professionals (vs. {31%} and [43%]). Overall, the balance between academics and external professionals seems to have remained fairly stable across the three surveys.

*Q6.3 (qualifications of trainers)* produced significantly different answers from EMT2017. Whereas in 2017, {75%} of programmes reported that half or more of their trainers held a formal qualification in tools, the figure in 2023 was 44%, marking a return to the 2012 level ([46%]). Conversely, 19% of programmes reported having no teachers with formal qualifications (vs. {6%} and [29%]), and a further 23% reported that most of their staff held no formal qualification. This dramatic apparent reversal of what the previous report identified as significant progress in professionalisation is not easy to explain, although it may be partly accounted for by the different constituency and high number of recently founded programmes that responded to EMT2023.

One of the 18 respondents to (optional) Q6.4 expressed doubt about what a 'formal qualification' in tools might be; others referred to PhDs, CPD and self-learning, and Trados Certification. As in 2017, several respondents also suggested that practical experience with tools in the language industry should be considered equivalent to a formal qualification.

The results of Q6.5 (experience of trainers) were similar to 2017. 94% of programmes reported that more than half of their tools teaching staff had five years' experience or more, as against {96%} in 2017 and just [70%] in 2012. As in 2017, no programme reported being entirely without such staff.

Among the 12 comments received in response to (optional) Q6.6, one noted that financial constraints had forced reduced use of external experts, another mentioned the use of PhD demonstrators, and a third the training of younger staff by those with greater experience. Several stressed the importance of industry experience among tools trainers on their programmes.

The next eight questions in EMT2023 related to student access to IT facilities and software licences. Q6.7 (ratio of students to workstations) revealed that the proportion of programmes with one computer for each student had risen again, from {78%} to 86%, and only 2 programmes reported a ratio less favourable than 1:2. Although several of the 11 respondents to Q6.8 mentioned computer labs, all but one pointed out that students (also) worked on their own devices, one remarking: "Online delivery. Why have computer labs when (our) students all have laptops?". This tendency to move away from conventional labs, already identified in EMT2017, has clearly accelerated, no doubt boosted by online delivery through the pandemic.
This could also help explain the slight regression seen in responses to *Q6.9 (out of class lab access),* where 71% of respondents confirmed that this is possible in their programmes, compared to {87%} in 2017 and [54%] in 2012. Of the 18 comments in response to Q6.10, several record extensive access hours through the week, and others identify local issues of managing lab access. However, more numerous are the comments reporting low student uptake: "It is a pity that they do not seem to need it anymore: they learned to use their own computers." One respondent simply states: "No labs", another: "We have no lab as such. They can access university computers." This is no doubt partly the result of the expansion of cloud tools, and also the fact that vendors of workstation-based tools tended to provide licences for students to use on their own devices during the pandemic.

This leads on to *Q6.11 (remote tools access),* to which 37% of respondents (up from {16%}) replied that all tools were accessible from off campus, 40% most tools (up from {20%}), and 19% some tools (as against {49%}), making a total of 96% (vs. {85%} and [58%]). Comments in response to Q6.12 seemed to confirm that off-campus access via licence servers had become less important with the expansion of cloud-based tools and the provision by vendors of licences for individual student use. One respondent noted that exercises had already been designed to be done from home.

*Q6.13 (integration of personal devices into tools classes)* confirmed a dramatic expansion in the use of personal devices and a steep decline in the sole use of computer labs. 32% of respondents said they relied on personal devices in one or more modules (vs. {11%} in 2017) and 63% said they were an optional extra in one or more modules (vs. {42%}), while just 5% (vs. {47%}) said that their classes used only lab facilities. The 14 free-text comments confirm that students often prefer to use their own laptop even in a computer lab, one respondent noting: "This sometimes causes technical/compatibility issues and more often causes ergonomic issues with things like screen size and access to network drives." Another wrote of the need to accommodate laptop access in mixed-mode classes for those students working remotely.

Q6.15 was simplified compared to 2017 to omit issues about server provision and concentrate only on technical support. Since the option to select more than one answer was removed this time in order to capture the primary means of support, results are not always directly comparable. 26% of respondents said their programmes had dedicated technical support (vs. {33%} in 2017 and [24%] in 2012), and a further 24% shared technical support with other specialist programmes, while the other 50% of programmes were reliant on generic institutional support. These figures probably confirm "the normalisation of the requirements for teaching translation software within universities" noted in 2017.

### 3.7 Innovative Practice

To *Q7.1* asking "Are there particular aspects of your translation programme that you believe represent elements of good / innovative / cutting-edge practice in the teaching of translation tools and technologies that you might wish to bring to the attention of others?", 30 respondents replied in the affirmative. For the full details, readers should consult the survey results,[i] but themes that emerged focus on teamwork and collaboration (for shared resources, project-based learning, usability and ergonomic projects, and for work on terminology, corpora, and glossary compilation), increased technical learning (using NLP techniques, Jupyter notebooks, app development), and mentoring from external professionals within a 'skills lab'.

### 3.8 Future Prospects

The aim of this section of EMT2023 was to capture the sentiment of respondents towards likely developments in the field of translation tools and technologies training in the medium term.

*Q8.1 (opportunities and challenges)* set out 13 opportunities and 5 possible threats, asking respondents to rank their probability according to a numerical scale. The expectation of increased student demand fell again, from [64%] in 2012 and {53%} in 2017, to just 32% this time, and of the creation of new study programmes from [50%] to {38%} and now 31%, though 44% of respondents still expected to develop at least one new collaborative programme (vs. [46%] and {45%}). On the other hand, the expectation of an expansion of the tools element of programmes was again sharply up, from [66%] and {78%} to 89%, while the expectation that new tools would be introduced remained stable at 79% (vs. [60%] and {80%}). Perhaps surprisingly, the expectation of new/alternative teaching methods such as e-learning was down at 39% (from {58%} and [50%]), though of course many such alternative methods are no longer 'new' since the pandemic.

Again surprisingly in the age of simplified web-based tools, 85% of respondents expected to see more complex technology (up from [64%] and {82%}), 71% expect tools to move (almost) completely to the cloud (vs. [58%] and {82%}, though the question in the previous surveys was less strongly worded). New for 2023, Q8.1.9 asked if it will no longer be necessary to provide workstations for students: 44% answered in the affirmative, while 32% remained neutral. 80% of respondents expected MT to become more important in the language industries (up from [58%] and {76%}). Another new question sought to refine this by asking whether PE (as a specific offering) would eventually become more frequent than human translation in certain areas: 84% thought this likely or very likely, while just 5% thought it unlikely and nobody saw it as very unlikely. As last time, 60% of respondents expected that more tools-qualified staff would become available, while expectations of an increase in the role of industry in programmes declined from 64% in both previous surveys, to just 50% this time.

In terms of perceived threats, 26% of respondents expected a lack of trained staff to hinder their operations (vs {22%} in 2017 and [36%] in 2012), and the fear of insufficient financial support also rose, from {42%} and [44%] to 52%. After a dramatic reduction from [26%] to {7%} in 2017, the perceived threat from a lack of IT facilities was back up to 21% this time, perhaps reflecting the number of respondents with recently established programmes. Maybe for the same reason, fears of a lack of technical support rose from {18%} in 2017 to 31%, and of a lack of support from government from {18%} to 34%.

Among the 19 respondents who commented on these future developments in Q8.2, several mentioned concerns about likely funding restrictions and lack of institutional support, particularly in the light of "falling student numbers post-Brexit", while one questioned the willingness of staff to engage with new technological developments. The uncertain impact of LLMs and tools such as GPT-4 was mentioned by two respondents as a concern, and another predicted a (possibly related) “shift towards more language engineering, [PM] and blended writing/translation skills”. One master’s programme is planning to move away from translation towards more advanced IT, in collaboration with a Computer Science department, and another respondent comments: “The profession is changing so much with the fast development of MT and AI that I'm not sure if there will be ‘translators’ as we know them now anymore, but different kinds of communication and language professionals using different kinds of tools. Which means I'm not sure if the university will see a need for a ‘translator training programme’ anymore.” However, the same person also expresses enthusiasm for “remote teaching exchange with other translation programmes”. A different geographical perspective comes from Japan, where according to one respondent the greater “emphasis on academic contents than practical contents” makes translation tools and technologies less of a priority for translation programmes than they have become elsewhere.

## 4. Limitations and Future Research

The EMT2023 survey achieved a lower response rate among EMT member programmes than the previous versions, but a wider (though non-systematic) distribution beyond the Network. The snapshot it presents can therefore not be regarded as rigorously representative of translator training around the world. Equally, it seems likely that programmes (both in and outside the EMT) that regarded themselves as more “advanced” in this area were the more likely to respond, so the picture that emerges almost certainly represents the practice of a substantial avant-garde, rather than an average. Recent research has seen increased interest in students’ perspectives on translation technology (see the Literature review section above). Although relevant and interesting to trainers, this aspect was not studied here as it would merit a separate research and publication project. There was no consistent and significant difference in responses between those from EMT members and non-member programmes.

## 5. Conclusions

Our main research question for this survey was to identify the state of the art for master’s level translation technology teaching, as represented by respondents at the time that the survey was conducted (between March and May 2023). In answering this, there are some significant and unambiguous findings to report. CAT tools remain central to technology teaching, with a move away from university labs to students’ own computers. Notable, however, is the responsiveness of programmes to innovations in translation technology: we found a major advance in the compulsory embeddedness of MT, especially from the PE and quality evaluation perspectives, and almost all respondents had or were planning to introduce LLMs and generative tools into their teaching. Another finding is the resilience and inventiveness of trainers facing the challenges of the Covid-19 pandemic, which in some cases has resulted in lasting improvements to course materials and flexible delivery. While the range of tools being taught has continued to expand, programmes seem to be consolidating their core offering, again no doubt influenced by the pandemic, around cloud-based software with cost-free academic access.

In answering the subquestion about how translation technology teaching has changed over the past ten years, interpreting the longitudinal developments revealed by the survey data is complicated by the wider geographical distribution of participants: for instance, whereas EMT2017 found few recently founded programmes in Europe, EMT2023 identified 24 (39% of responses) launched since 2010. We can report an increase in the embedding of professional contexts and workflows associated with translation technology. 70% of programmes (up from {55%}) reported that tools were now compulsory in practical translation classes and 63% (a significant rise from {45%}) that they are mandatory in translation theory modules. Tools remain a compulsory element of over 60% of modules on the translation profession and skills lab-type modules. Information mining increased again to 85%, reinforcing the central role played by data-gathering skills. The top technology-related themes, as in both previous surveys, were tools as part of a translation-related project (82%, up from {75%} and [58%]) and tools in professional roles and workflows. Generic file management and data security skills have increased in perceived importance, and legal and ethical issues related to translation data have also become more prominent.

In terms of course delivery the shift away from conventional labs identified in EMT2017 has accelerated markedly, no doubt partly driven by the pandemic, accompanied by a dramatic expansion in the use of students' personal devices. While expectations in 2017 that tools would move to the cloud and MT would become more important in the language industries have been realised, the decline in confidence that the sector would expand with new programmes and more students has continued this time, although expectations that the tools element of existing programmes would expand were again sharply up.

Finally, looking at how respondents see programmes changing in the short- to medium-term, 84% of respondents thought it likely or very likely that PE (as a specific offering) will eventually become more frequent than human translation in certain areas, while none considered it very unlikely. The expansion of technology within translation programmes seems set to increase, with reports of NLP techniques, Jupyter notebooks, and app development being added to programmes. There appears to be some uncertainty about the future due to sociopolitical considerations, such a Brexit and a changing political climate that bring concerns about future financial, institutional and political support for translation programmes to the fore. And while programmes showed an agility in response to the launch of LLMs and generative tools, some respondents are looking to make substantial changes to their programmes tailored to the use of these tools.

## Acknowledgements

The authors would like to thank the survey participants, the anonymous reviewers of drafts of this text, and colleagues at the EMT Network and the DG-T.

---

[i] The anonymised full dataset of responses may be downloaded from https://github.com/jossm0/EMT2023/blob/main/TTT%20Survey%202023%20anon.xlsx.

**Appendix: Survey Questions**

**European Master's in Translation Network: Translation Tools and Technologies Survey, 2023**

This is the third in a regular cycle of surveys designed to monitor the evolution of training in translation tools and technologies in master's level translation programmes around Europe, but now we would like to extend this to programmes worldwide. The EMT Network and Board would be most grateful if you could take the time to complete it on behalf of your translator training programme(s). The results will be widely disseminated and are expected to inform training policy and the future design of programmes. If you are not the person best placed to do this, please forward the survey link to the person who is (but please ensure you only make one return for each programme).

It's important to note that no programme will be identified in publication of the results and there is no judgement of programmes inferred - quite the opposite: we hope that the results might help to encourage additional resources and share best practices.

If you have more than one programme, we request that you fill out the form once for each. This does mean more work, but will allow more accurate classification of results, as well as making it clearer how to respond to questions where the answer might be different for each programme.

The first survey was conducted in 2012 as part of the EU-funded OPTIMALE project. The second survey was run by the Tools and Technologies Working Group in the EMT Network in 2017. Please see an article reporting on this survey (in English) here.

For the purposes of the present survey, 'translation tools and technologies' refers to software applications and systems specifically designed to assist in the translation process, or to perform translation automatically, which trainee translators must learn to use. For comparability with the 2012 and 2017 surveys, many of the original questions have been retained, though in some cases with updates (e.g. the list of tools in Section 4). Some new questions have been added for 2023: these are preceded by [New].

In order to submit your completed survey, it is necessary to answer all the compulsory questions on nine survey pages. We estimate that this will take a minimum of 20 minutes, and we thank you in advance for taking the time to do this. It is possible to save a part-finished survey and return to it later.

DEADLINE FOR COMPLETION: Monday 15th May 2023, 23.59 CET

Joss Moorkens, Andrew Rothwell, and Tomáš Svoboda

EMT Tools & Technologies Working Group

If you have any questions, please contact one of us at joss.moorkens@dcu.ie, a.j.rothwell@swansea.ac.uk, or tomas.svoboda@ff.cuni.cz

Sign in to Google to save your progress. Learn more

* Indicates required question

Email*

Your email address

This is a required question

---

**1: Administrative details**

**1.1** Name of the university and department/institute offering the master's programme:
**1.2** Programme title (in English, please):

**1.3** Programme website address:

**1.4** Network memberships:* (Y/N)

Is the programme a current EMT member?

Was the programme an EMT member in 2017?

Was the programme an EMT member in 2012?

[New] **1.4.1** Is your programme a member of any other national or international network? If so, which one(s)?

**1.5** ECTS value of the programme:*

**1.6** (Optional) If you selected 'other', please specify here:

**1.7** Year the programme was launched (if the title has changed, please give the original year):*

**1.8** Average number of students (full and part time) per year in the last 5 years:*

---

**2: Overall approach to tools training**

**2.1** Mark all relevant options* (Y/N)

We currently teach and assess translation tools in theory

We currently teach and assess translation tools in practice

**2.2** What is the approximate minimum percentage of (compulsory) study time (of the whole programme) that a student must devote to translation tools?*

**2.3** (Optional) If more than 25%, please explain:

**2.4** What is the approximate maximum percentage of (compulsory + optional) study time (of the whole programme) that a student can devote to translation tools?*

**2.5** (Optional) If more than 50%, please explain:

**2.6** How many COMPULSORY credits are mainly devoted to tools and technologies (i.e. where learning technologies are the main focus of the module)?*

**2.7** (Optional) Please name/describe compulsory modules/elements:

**2.8** In addition, how many OPTIONAL credits are mainly devoted to tools and technologies (i.e. where learning technologies are the main focus of the module)?*

**2.8.1** (Optional) Please name/describe optional modules/elements:

**2.9** Your approach to tools training:* (Y/N)

**2.9.1** We teach and assess open-access tools

**2.9.2** We teach and assess commercial tools

**2.9.3** We teach tools from the perspective of translators

**2.9.4** We teach tools from the perspective of project managers (e.g. management of suppliers and TMs etc.)

**2.9.5** We teach tools from the perspective of translation companies (e.g. client portals)

**2.9.6** Students with different language combinations work together in tools training exercises

---

**3: Types of training activities**

**3.1** Please indicate which activities involving translation--related technologies you teach, whether they are compulsory or optional, and whether they play a major or minor role in your programme (e.g. an optional activity may be a major component of the programme for students who take it):*

(Not taught / Compulsory – major / Compulsory – minor/ Optional – major / Optional – minor)

**3.1.1** Information mining - search strategies

**3.1.2** Information mining - evaluation of resources

**3.1.3** Text and/or corpus analysis using concordancers etc.

**3.1.4** Use of term bases

**3.1.5** Computerised terminology extraction

**3.1.6** Translation Memory use

**3.1.7** Translation Memory construction (alignment and/or import)

**3.1.8** Use of shared Translation Memories

**3.1.9** Use of shared termbases

**3.1.10** Machine Translation used / post-edited in a CAT tool interface

**3.1.11** Machine Translation used / post-edited outside a CAT tool

[New] **3.1.12** Use of interactive/adaptive MT?

**3.1.13** Construction/training of MT engines

**3.1.14** Human evaluation of MT output

**3.1.15** Automatic evaluation of MT output using string-based metrics (e.g. BLEU)

[New] **3.1.16** Automatic evaluation of MT output using pre-trained metrics (e.g. COMET)

**3.1.17** Website localization

**3.1.18** Software (not games) localization

**3.1.19** Games localization

**3.1.20** Multimedia translation (desktop subtitling)

**3.1.21** Multimedia translation (cloud subtitling)

**3.1.22** Multimedia translation (dubbing / voiceover)

**3.1.23** Multimedia translation/media accessibility (e.g. audio description)

**3.1.24** Translation management systems

**3.1.25** Quality Assurance features of CAT tools

**3.1.26** Quality Assurance - standalone tools

**3.1.27** Web editing

**3.1.28** Desktop publishing

**3.1.29** Optical Character Recognition (OCR)

**3.1.30** Automatic Speech Recognition (ASR)

**3.2** Please comment on the overall mix of technology-based activities in your programme, including any not listed above:

**3.3** Please identify the role of tools in modules / course units not specifically devoted to them:*

(Not applicable / Compulsory – major / Compulsory – minor / Optional – major / Optional – minor)

**3.3.1** Practical translation classes

**3.3.2** Introduction to the translation profession / market

**3.3.3** Translation theory

**3.3.4** Skills lab / simulated translation company

**3.3.5** Internship

**3.3.6** Extended translation project

**3.3.7** Dissertation

**3.4** (Optional) Comments on question 3.3:

---

**4: Translation software taught and licences held**

**4.1.1** Please indicate which CAT, terminology, or localisation software you teach, whether each package is compulsory or optional, and the approximate ratio of students to licences that you hold (e.g. ≤1 means that you have one licence per student or better, 1.1 - 2.9 that there are between 1 and 3 students for every licence, and ≥3 that there are three or more students for every licence). We ask about MT and AVT below.*

(Compulsory ≤1 / Compulsory 1.1 - 2.9 / Compulsory ≥3 / Optional ≤1 / Optional 1.1 - 2.9 / Optional ≥3 / Not taught)

Across

Alchemy Catalyst

Atril Déjà Vu

Fluency Now

Lilt

Lingobit Localizer

[New] Lokalise

Matecat

memoQ

MetaTexis

MultiTerm

MultiTrans

OmegaT

Passolo

Phrase TMS (formerly Memsource)

[New} Plunet

Similis

[New] Smartcat

[New] Smartling

Star Transit

Termstar

Trados Studio

Wordbee

Wordfast Anywhere

Wordfast Classic

Wordfast Pro

XTM

[New] XTRF

**4.1.2** Please indicate which MT tools you teach, whether each package is compulsory or optional, and the approximate ratio of students to licences that you hold (e.g. ≤1 means that you have one licence per student or better, 1.1 - 2.9 that there are between 1 and 3 students for every licence, and ≥3 that there are three or more students for every licence)*

(Compulsory ≤1 / Compulsory 1.1 - 2.9 / Compulsory ≥3 / Optional ≤1 / Optional 1.1 - 2.9 / Optional ≥3 / Not taught)

[New] DeepL

[New] eTranslation

GoogleTranslate

KantanAI

Microsoft (Bing) Translator

Opus-MT

Systran

**4.1.3** Please indicate which audiovisual translation software you teach, whether each package is compulsory or optional, and the approximate ratio of students to licences that you hold (e.g. ≤1 means that you have one licence per student or better, 1.1 - 2.9 that there are between 1 and 3 students for every licence, and ≥3 that there are three or more students for every licence)*

(Compulsory ≤1 / Compulsory 1.1 - 2.9 / Compulsory ≥3 / Optional ≤1 / Optional 1.1 - 2.9 / Optional ≥3 / Not taught)

[New] Aegisub

[New] EZTitles

[New] Fab Subtitler

[New] MateSub

[New] Ooona Tools

[New] SPOT

[New] Subtitle Edit

[New] Subtitle Workshop

[New] Swift

[New] Trados Subtitle App

[New] WinCaps

[New] Zoo Dubs

[New] Zoo Subs

**4.2** Please describe your tools selection and licence holding strategy or tools taught not mentioned above, including any relevant constraints and limitations:*

[New] **4.3** Are there any tools or technologies that you would like to teach but are unable to? If so, why not?*

[New] **4.4** Do you teach integration of MT into CAT tools (e.g. by API plugin)?* (Y/N)

[New] **4.4.1** If you answered 'yes' above, please provide some details.

[New] **4.5** Do you use paid versions of free online MT tools? * (Y/N)

[New] **4.5.1** If you answered 'yes' above, please provide some details.

[New] **4.6** Do you teach about ChatGPT or another text generator tool for writing or translation? If so, please provide some details.

---

**5: Tools teaching and assessment themes**

**5.1** Please indicate which of the following themes your programme teaches, and rate their importance numerically.*

(1=very important, 2=important, 3=neutral, 4= not very important, 5=not important at all, [Not taught])

**5.1.1** History and development of translation tools

**5.1.2** Theory and principles of translation tools

**5.1.3** Generic file management and data security skills

**5.1.4** Advanced Office skills (e.g. macros, mailmerge)

**5.1.5** Tools in and for themselves (i.e. as software packages)

**5.1.6** Tools as part of a translation-related project

**5.1.7** Professional roles and workflows

[New] **5.1.8** Ergonomic issues with post-editing of Machine Translation

**5.1.9** Translation industry structure and future development

**5.1.10** Industry standards (e.g. ISO 17100)

**5.1.11** Legal issues (e.g. around TM/language data ownership)

**5.1.12** Crowd-sourcing and collaborative translation

**5.1.13** Volunteer translation

[New] **5.1.14** Ethics of translation technology

[New] **5.1.15** Large language models (e.g. GPT3, ChatGPT)

**5.1.16** Other (please explain in 5.2)

**5.2** Please comment on your answers in 5.1:

**5.3** Please indicate which tools teaching and assessment strategies your programme uses, and rate their importance numerically.*

(1=very important, 2=important, 3=neutral, 4=not very important, 5=not important at all, [Not used])

**5.3.1** Learning through staff lecture or demonstration

**5.3.2** Autonomous learning from manuals and Help systems

**5.3.3** E-learning (online delivery)

**5.3.4** Blended learning

**5.3.5** Learning through individual work

**5.3.6** Learning through team or group work

**5.3.7** Assessment by individual work

**5.3.8** Assessment by group or team work

**5.3.9** Assessment by practical tools-based task

**5.3.10** Assessment by analytical and descriptive report

**5.3.11** Assessment by general essay

**5.3.12** Assessment by tool vendor certification tests

**5.3.13** Other (please explain in 5.4)

**5.4** Please comment on your answers in 5.3:

[New] **5.5** Did the necessity of remote delivery and collaboration during the COVID-19 pandemic have an influence on the i) content ii) delivery iii) assessment of your programme? If yes, please describe in more detail below.

[New] **5.6** Have you retained any of these changes post pandemic?

---

**6: Staff training, IT facilities and technical support**

**6.1** Who teaches tools and technologies in your programme?*

**6.2** (Optional) Comments on teaching staff

**6.3** What qualifications do your tools teachers have?*

**6.4** (Optional) Comments on qualifications

**6.5** How much experience do tools teachers have?*

**6.6** (Optional) Comments on experience, both as teachers and users of translation technology

**6.7** Facilities: what is the ratio of students in tools classes to tools-equipped networked workstations?

**6.8** (Optional) Comments on students / workstations ratio

**6.9** Can students access labs outside class time?* (Y/N)

**6.10** (Optional) Comments on lab access

**6.11** Can students access software tools remotely?*

**6.12** (Optional) Comments on remote access

**6.13** Are students' personal devices integrated into your tools classes?*

**6.14** (Optional) Please comment on your answers to 6.13.

**6.15** Technical support.*

**6.16** (Optional) Comments on technical support

---

**7: Innovative practice**

**7.1** Are there particular aspects of your translation programme that you believe represent elements of good / innovative / cutting-edge practice in the teaching of translation tools and technologies that you might wish to bring to the attention of others?

**7.2** (Optional) Please describe any aspects of your programme that you regard as examples of innovative practice

**8: Future prospects**

**8.1** Looking ahead over the next five years, how likely is your programme to experience the following opportunities and challenges? *

(1=very likely, 2=likely, 3=neutral, 4=unlikely, 5=very unlikely):

**8.1.1** Student demand for our Translation programme(s) will increase

**8.1.2** We expect to develop one or more new translation-related programme(s)

**8.1.3** We expect the translation technology element of our programme(s) to expand

**8.1.4** We expect to introduce training in different types of translation tools

**8.1.5** We expect to develop one or more collaborative programme(s) with other institutions

**8.1.6** We expect to introduce / expand alternative teaching methods (e.g. e-learning)

**8.1.7** Translation technology will become more complex and diverse

**8.1.8** Translation tools will migrate (almost) completely from local installations to the Cloud

[New] **8.1.9** We will no longer need to provide workstations for students as they will use their own devices

**8.1.10** Fully automatic translation (MT) will become more important in the industry

[New] **8.1.11** MT post-editing (as a specific service offering) will eventually become more frequent than human translation in certain areas

**8.1.12** Newly qualified staff with translation technology training will become available

**8.1.13** Industry involvement with translation training programmes will increase

**8.1.14** Our future developments will be limited by lack of trained staff

**8.1.15** Our future developments will be limited by lack of money

**8.1.16** Our future developments will be limited by lack of IT facilities

**8.1.17** Our future developments will be limited by lack of technical support

**8.1.18** Our future developments will be limited by lack of official (institutional/governmental) support

**8.2** (Optional) Please comment on how you see future developments in the translation technology element of your programme(s):

---

**9: Respondent contact details**

**9.1** Your title and full name:*

**9.2** Your institutional/departmental affiliation, and full address:*

**9.3** Your role (e.g. Programme Director, Tools trainer...)*

END